\documentclass[prx,superscriptaddress,twocolumn]{revtex4}
\usepackage{amsmath, amsthm, amssymb}
\usepackage{mathtools}
\usepackage[pdftex]{graphicx}
\usepackage{times}
\usepackage{subfigure}
\usepackage{hhline}
\usepackage{mathrsfs}
\usepackage{color,soul}
\usepackage[normalem]{ulem}
\usepackage{natbib}
\usepackage{bbold}
\usepackage{placeins}
\usepackage{tcolorbox}
\usepackage{braket}
\usepackage{soul,xcolor}
\usepackage{tabu}
\usepackage{array}
\usepackage{longtable}
\usepackage{multirow}
\usepackage{tikz}
\usetikzlibrary{decorations.pathmorphing,calc}
\usetikzlibrary{shapes.geometric,positioning, arrows.meta,calc}
\tikzstyle{box} = [draw, fill=blue!10, text centered, rectangle, minimum width=2cm, minimum height=1cm]
\tikzstyle{arrow} = [thick,->,>=stealth]
\usepackage{enumitem}
\usepackage[english]{babel}

\usepackage{bm}

\usepackage[utf8]{inputenc}

\definecolor{myurlcolor}{rgb}{0,0,0.7}
\definecolor{myrefcolor}{rgb}{0.8,0,0}
\usepackage[unicode=true,pdfusetitle, bookmarks=false,bookmarksnumbered=false,
bookmarksopen=false, breaklinks=false,pdfborder={0 0 0},backref=false,
colorlinks=true, linkcolor=myrefcolor,citecolor=myurlcolor,urlcolor=myurlcolor]{hyperref}

\DeclareGraphicsExtensions{.png,.pdf,.eps,.jpg}
\graphicspath{ {./figs/} }
\newcommand{\eqnref}[1]{Eq.~(\ref{#1})}

\newcommand{\figref}[1]{Fig.~\ref{#1}}
\newcommand{\tabref}[1]{Table~\ref{#1}}
\newcommand{\secref}[1]{Sec.~\ref{#1}}
\newcommand{\appref}[1]{App.~\ref{#1}}

\renewcommand\Re{\operatorname{Re}}
\renewcommand\Im{\operatorname{Im}}

\newcommand{\Tr}{\operatorname{Tr}}
\newcommand{\order}[1]{\mathcal{O}(#1)}
\newcommand{\rank}{\operatorname{rank}}

\newcolumntype{M}[1]{>{\centering\arraybackslash}m{#1}}
\newcolumntype{N}{@{}m{0pt}@{}}

\newcommand{\mrm}[1]{\mathrm{#1}}

\renewcommand{\d}{\mrm{d}}
\renewcommand{\i}{\mrm{i}}

\newcommand{\T}{{\mrm{T}}}

\newcommand{\qmean}[1]{{\langle#1\rangle}}

\newcommand{\mat}[1]{{\textbf{#1}}}

\newcommand{\qvec}[1]{\hat{\bm{#1}}}
\newcommand{\id}{\mat{I}}   

\newcommand{\smat}[1]{{\mathsf{#1}}}
\newcommand{\sformletter}{J}
\newcommand{\sform}{{\smat{\sformletter}}} 
\newcommand{\stransf}[1]{{\mathcal{\sformletter}\!\left\{#1\right\}}}

\newcommand{\svec}[1]{{\smat{#1}}}
\newcommand{\qsvec}[1]{{\hat{\smat{#1}}}}
\newcommand{\Id}{\smat{I}}	 

\newcommand{\dprime}{{\prime \prime}}

\newcommand{\X}{{\smat{X}}}
\newcommand{\A}{{\smat{A}}}

\newcommand{\beq}{\begin{equation}}
\newcommand{\eeq}{\end{equation}}

\newcommand{\bpm}{\begin{pmatrix}}
\newcommand{\epm}{\end{pmatrix}}

\newcommand{\ben}{\begin{enumerate}}
\newcommand{\een}{\end{enumerate}}

\newcommand{\bitm}{\begin{itemize}}
\newcommand{\eitm}{\end{itemize}}

	\date{\today}
	 
\begin{document} 
	\title{Enhanced quantum sensing with hybrid exceptional--diabolic singularities}

	\author{Javid Naikoo}
	\email{javid.naikoo@amu.edu.pl}
	
	\author{Ravindra W. Chhajlany}

		\author{Adam Miranowicz}
	\affiliation{Institute of Spintronics and Quantum Information, Faculty of Physics and Astronomy, Adam Mickiewicz University, 61-614 Pozna\'{n}, Poland}

	\begin{abstract}
We report an enhanced sensitivity for detecting linear perturbations near hybrid (doubly degenerated) exceptional-diabolic (HED) singular points in a four mode bosonic system. The sensitivity enhancement is attributed to a singular response function, with the pole order determining  the scaling of estimation error. At  HED singular points, the error scaling exhibits a twofold improvement over non-HED singular points. The ultimate bound on estimation error is derived via quantum Fisher information, with heterodyne detection identified as the measurement achieving this optimal scaling.
	\end{abstract}
	
	\maketitle
	\section{Introduction}
 
	The study of non-Hermitian systems, which feature complex eigenvalues, has gained significant attention due to their intriguing and unconventional properties. A key phenomenon in such systems is the presence of \textit{exceptional points} (EPs)—\emph{singularities} in the complex energy plane where both eigenvalues and eigenvectors coalesce. First studied in physical context in \cite{Kato1966}, EPs arise in parameter-dependent eigenvalue problems and are encountered in a wide range of physical systems, such as optics, mechanics, quantum phase transitions, and quantum chaos~\cite{Ozdemit2019,ElGanainy2018,Parto2020,Heiss2012}. In contrast, \textit{diabolic points} (DPs), introduced in \cite{Berry1984}, arise in Hermitian systems as instances of eigenvalue degeneracies where eigenvectors remain orthogonal. These points are closely associated with geometric phases and Berry's curvature. Remarkably, certain non-Hermitian systems can exhibit both EPs and DPs, with their intersections giving rise to \textit{hybrid exceptional-diabolic} (HED) points. These HED points blend the distinct features of both singularities  and  offers valuable insights into the interplay between Hermitian and non-Hermitian physics unlocking potential applications across diverse physical domains~\cite{Perina2022Quantum,arkhipov2023dynamically,Thapliyal2024Multiple,Perina2024Multiple}.

	Recent advances in quantum sensing have highlighted the transformative potential of non-Hermitian systems in achieving ultra-sensitive measurements, especially through the use of exceptional point singularities, where small perturbations lead to disproportionately large responses. This property can dramatically amplify sensitivity, offering a pathway to high-precision measurements that would be infeasible in Hermitian frameworks~\cite{WiersigSensing14, Heiss2004, Miri2019, Ravi2019}. 
	A system near an EP, when perturbed, displays eigenvalue splitting that scales as the $n$th root of the perturbation size ($\sim \!\sqrt[n]{\epsilon}$), yielding a steep sensitivity that can surpass what is achievable in Hermitian settings where perturbations result in linear or quadratic responses~\cite{Wiersig20review}. Experimental work with optical resonators has validated this principle, demonstrating that under carefully controlled conditions, non-Hermitian systems can achieve extremely sensitive measurements~\cite{Chen2017, Hodaei2017}. However, under real-world noise conditions, this sensitivity is typically limited by quantum noise, which can diminish the response and limit the achievable gains~\cite{Langbein2018,Lau2018,Mortensen2018,Wolff2019,Zhang2019,Chen2019,Naikoo2023,Loughlin2024,Wiersig2020}.
	
	In particular,  \cite{Langbein2018,Chen2019} argue convincingly that EP sensors do not provide any fundamental improvement in signal-to-noise ratio (SNR), challenging the notion that exceptional points inherently enhance sensor performance. Building on this, \cite{Lau2018}  highlights that enhanced sensitivity is primarily driven by nonreciprocity rather than the proximity to an exceptional point. This study establishes fundamental bounds on signal power and SNR for two-mode sensors, demonstrating that while gain is necessary for enhanced signal power, it does not rely on being near an EP. Furthermore, it shows that when fluctuation effects are considered, the SNR of any reciprocal system is fundamentally constrained, regardless of its use of EPs. These insights critically address the limitations of EP-based sensing, shifting the focus from EP proximity to more significant factors such as nonreciprocal interactions and gain in the system dynamics.
	
	Instead of relying on eigenmode coalescence at EPs,  singularity-tuned systems take advantage of internal parameter tuning that amplifies response to perturbations -- a sensitivity boost that differs fundamentally from that of EP-based systems~\cite{Zhang2019,Naikoo2023}. This divergence is related to proximity to a dynamical phase transition, which has recently been recognized as a valuable resource in quantum sensing. Unlike the more conventional amplification methods, this approach benefits from the unique properties of singularities in the non-Hermitian space, creating a form of criticality-enhanced sensitivity that is robust under a wider range of noise conditions than EP-based systems~\cite{Macieszczak2016, Fernandez2017, Wald2020, chu_dynamic_2021, wu_criticality-enhanced_2021}.
	
	In \cite{Naikoo2023}, it was demonstrated that in a two-mode bosonic system, the error \(\delta \theta\) in estimating a parameter \(\theta\) scales as \(\delta \theta \propto \theta^s\), where \(s\) corresponds to the order of the pole in the singular response function. Notably, when the parameter affects the common frequency and the singularity arises from satisfying the EP condition, the error exhibits  \emph{quadratic} scaling. This work extends the previous analysis by generalizing it to a four-mode cavity system featuring both a second-order EP and a second-order DP. We demonstrate that quadratic scaling persists at the second-order EP, even when an orthogonal subspace emerges due to the second-order DP. In the system under consideration, satisfying both the EP condition and the singularity condition of the dynamical generator necessitates meeting the diabolic condition. As a result, enhanced precision is observed specifically at HED points.

	The structure of this paper is as follows: Section~\ref{sec:Fisher} provides a concise overview of the classical Fisher information and its quantum counterpart, the quantum Fisher information, emphasizing their critical roles in estimation theory. Section~\ref{sec:Model} presents a comprehensive description of a sensor model, including the mathematical framework for Gaussian estimation of linear perturbations within this context. The primary results of this study are discussed in Sec.~\ref{sec:Analysis}, where we analyze the sensing advantages at HEP singularities. Finally, the conclusions are summarized in Sec.~\ref{sec:Conclusion}.

\section{Fundamental Bounds on Parameter Estimation: The Role of Fisher Information}\label{sec:Fisher}

The Fisher information and the Cram\'{e}r-Rao bound are foundational concepts in estimation theory, quantifying the limits of precision in parameter estimation. These concepts are pivotal in both classical and quantum domains, offering insight into the optimal accuracy achievable under the constraints of measurement and inherent noise.

\subsection{Classical Fisher information and the Cram\'{e}r-Rao bound}

In classical statistics, Fisher information provides a measure of the sensitivity of a probability distribution to the changes in a parameter. For a probability distribution $ p(x|\theta) $ that depends on an unknown parameter $ \theta $, the Fisher information $F_{\theta}$ is defined as
\begin{equation}
	F_{\theta} = \mathbb{E}\left[ \left( \frac{\partial}{\partial \theta} \log p(x|\theta) \right)^2 \right],
\end{equation}
where $ \mathbb{E} $ denotes the expectation over the variable $ x $ sampled from $ p(x|\theta) $. Intuitively, a higher Fisher information value indicates that small changes in $ \theta $ lead to larger variations in the probability distribution, suggesting that $ \theta $ can be estimated with higher precision.

The Cram\'{e}r-Rao bound establishes a lower bound on the variance of any unbiased estimator $ \hat{\theta} $ of the parameter $ \theta $, providing a measure of the best possible accuracy achievable in estimating $ \theta $ under unbiased conditions. For an unbiased estimator, the classical Cram\'{e}r-Rao  bound  is given by:
\begin{equation}
	\delta^2(\hat{\theta}) \geq \frac{1}{F_{\theta}},
\end{equation}
where $ \delta^2(\hat{\theta}) $ denotes the variance of $ \hat{\theta} $. This inequality implies that the inverse of the Fisher information is the lowest achievable variance for any unbiased estimator, setting a theoretical limit on estimation precision.

\subsection{Quantum Fisher information and the quantum Cram\'{e}r-Rao bound}
In quantum mechanics, the Fisher information concept extends to account for the intrinsic probabilistic nature of quantum measurements. Quantum Fisher information (QFI) characterizes how sensitively the state of a quantum system changes with respect to a parameter $ \theta $, taking into account both the probabilistic nature of measurement outcomes and the underlying quantum state dynamics.

Consider a quantum state $ \rho(\theta) $ that depends on the parameter $ \theta $. The QFI $ \mathcal{F}_{\theta} $ for this state is defined as
\begin{equation}
	\mathcal{F}_{\theta}  = \text{Tr} \left[ \rho(\theta) L_\theta^2 \right],
\end{equation}
where $ L_\theta $ is the symmetric logarithmic derivative (SLD) operator, implicitly defined by the equation:
\begin{equation}
	\frac{\partial \rho(\theta)}{\partial \theta} = \frac{1}{2} \bigg[ L_\theta \rho(\theta) + \rho(\theta) L_\theta \bigg].
\end{equation}
The SLD, $ L_\theta $, serves as a quantum analog to the derivative of the log-likelihood function in the classical case, capturing how changes in $ \theta $ affect the quantum state. The QFI is crucial in fields such as quantum metrology, where it enables the design of measurement protocols that maximize parameter estimation precision under quantum mechanical constraints.

The precision limits imposed by the QFI are formalized by the \textit{quantum Cram\'{e}r-Rao bound}, which sets a fundamental limit on the variance of any unbiased estimator $ \hat{\theta} $ for the parameter $ \theta $. The quantum Cram\'{e}r-Rao bound states that $\delta^2(\hat{\theta}) \geq \frac{1}{\mathcal{F}_{\theta}}$ and satisfies

\begin{equation}
	\delta^2(\hat{\theta}) \ge \frac{1}{F_{\theta}} \geq \frac{1}{\mathcal{F}_{\theta}}.
\end{equation}
This bound indicates that the inverse of the QFI is the minimum achievable variance, setting an ultimate limit on precision that is attainable by optimizing both  measurement and  quantum state preparation.

In practical terms, the quantum Cram\'{e}r-Rao bound plays a central role in quantum sensing and metrology, guiding the development of quantum-enhanced measurement strategies that leverage entanglement, coherence, and other uniquely quantum properties to surpass classical precision limits. By carefully engineering quantum states and measurements to maximize $ \mathcal{F}_{\theta} $, it is possible to achieve higher sensitivity in estimating parameters such as phase, frequency, and other physical quantities of interest in quantum systems.

	\section{Sensor model: A bosonic system with hybrid exceptonal and diabolic points}\label{sec:Model}

\begin{figure}[h!]
\includegraphics[width=\linewidth]{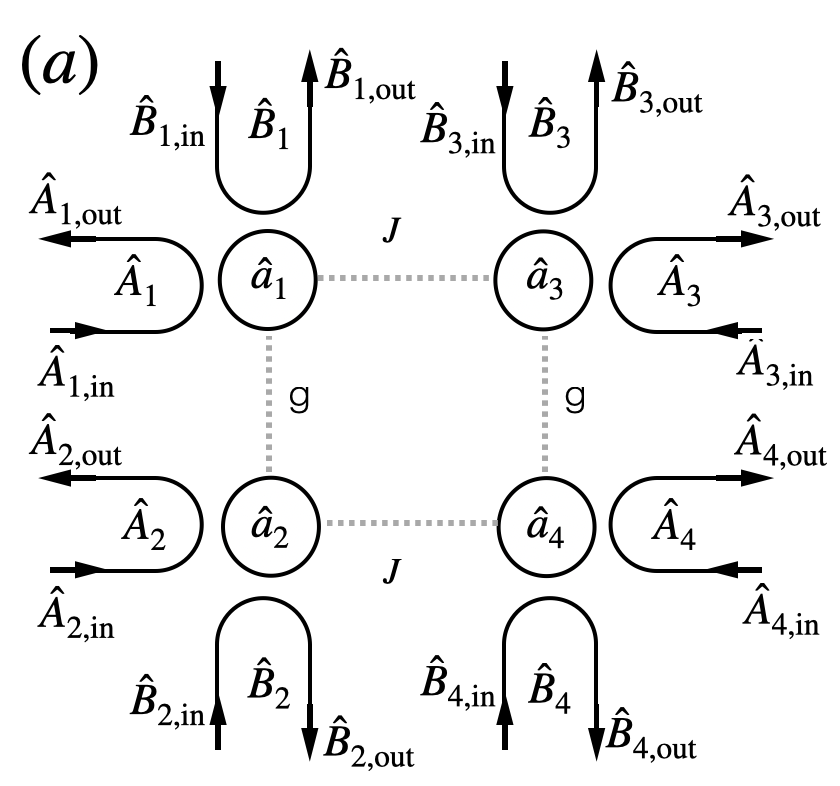}
\includegraphics[width=\linewidth]{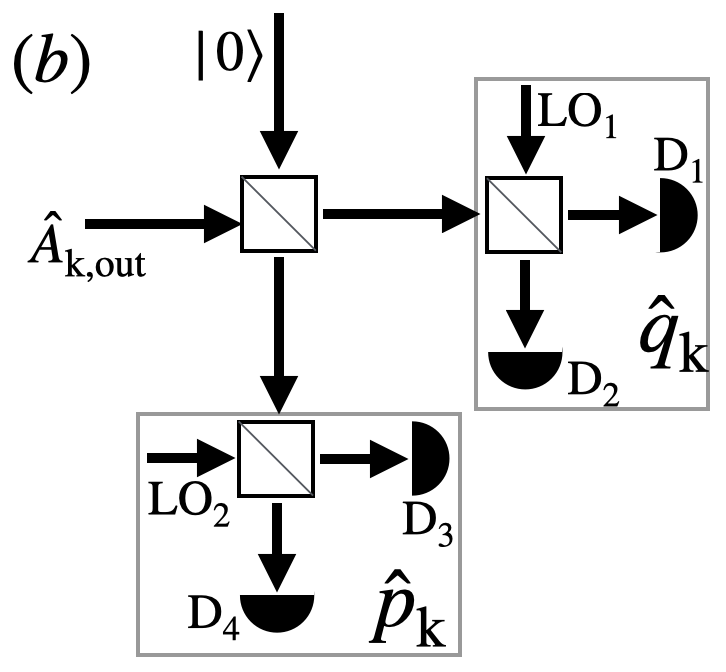}
	\caption{(a) \textbf{Sensor model} of four bosonic modes $\hat{a}_\mrm{k}$ interacting with \emph{inaccessible} bath modes $\hat{B}_\mrm{k}$ through dissipative (for $\mrm{k} \!\! =\!\! 1,3$)  and amplification (for $\mrm{k} \!\!=\!\! 2,4$) interactions. All the four cavity modes are probed by \emph{accessible} (Gaussian) input field $\hat{A}_\mrm{k}$ via dissipative interaction. Using input-output formalism, the accessible and the inaccessible field modes are expressed in terms of the input and output fields ($\hat{A}_\mrm{k,in/out}$ and $\hat{B}_\mrm{k,in/out}$ ) as defined in \eqnref{eq:InputOutputField}. (b) \textbf{Heterodyne detection}, which is equivalent to double \emph{homodyne} detection, is performed on the outputs from the accessible modes i.e. $\hat{A}_{k, \mathrm{out}}$, by combining them with the vacuum $|0\rangle$ on a balanced beam splitter. The resulting fields are then mixed with local oscillators ($\mathrm{LO}_1$ and $\mathrm{LO}_2$) on two additional balanced beam splitters. The outputs from these are measured by detectors ($\mathrm{D}_\mrm{k}$) to determine the quadratures $\hat{p}_\mrm{k}$ and $\hat{q}_\mrm{k}$.
	}
	\label{fig:model4modes}
\end{figure}
	Let us consider a system of four bosonic modes $\hat{a}_{\ell}$ $\ell \!\!=\!\! 1,\dots,4$, described by the following Hamiltonian
	\begin{align}
		\hat{H}_S &= \sum\limits_{\ell = 1}^{4} \omega_k \hat{a}_{k}^{\dagger} \hat{a}_{k} + g (\hat{a}_{1}^{\dagger} \hat{a}_{2}  + \hat{a}_{3}^{\dagger} \hat{a}_{4}  + \text{h.c.}) \nonumber \\&+ J (\hat{a}_{1}^{\dagger} \hat{a}_{3} + \hat{a}_{2}^{\dagger} \hat{a}_{4} + \rm h.c.).
	\end{align}
	We assume that the odd-numbered modes, $\hat{a}_{1}, \hat{a}_{3}$ and the even-numbered modes, 
	$\hat{a}_{2}, \hat{a}_{4}$ experience loss and gain, respectively. This loss-gain dynamics can be modeled through the following interactions:
	\begin{align}
H_\mrm{int, odd} &= \sum\limits_\mrm{\ell = odd}  \int_{-\infty}^{\infty} d \omega^\prime ~ \sqrt{\frac{\eta_{\ell} (\omega^\prime)}{2\pi}} [  \hat{a}_{\ell}  \hat{B}_{\ell}^\dagger(\omega^\prime)  + {\rm h.c.} ],\\
H_\mrm{int, even} &= \sum\limits_\mrm{\ell = even}  \int_{-\infty}^{\infty} d \omega^\prime ~ \sqrt{\frac{\eta_{\ell} (\omega^\prime)}{2\pi}} [  \hat{a}_{\ell}^{\dagger} \hat{B}_{\ell}^\dagger(\omega^\prime) + {\rm h.c.} ].
	\end{align}
	Here, the bath mode operators $\hat{B}_{1(3)}$ and $\hat{B}_{2(4)}$ are responsible for the \emph{intrinsic} loss and gain in the cavity modes  $\hat{a}_{1(3)}$ and the mode $\hat{a}_{2(4)}$, respectively. Further, we allow each cavity mode $\hat{a}_{\ell}$   to be  probed by a controllable field $\hat{A}_{\ell}$ and such probing  introduces further loss to each cavity mode $\hat{a}_{\ell}$ modeled by the following interaction
	\begin{equation}
	H_\mrm{int, probe}	= \sum\limits_{\ell = 1}^{4}    \int_{-\infty}^{\infty} d \omega^\prime ~ \sqrt{\frac{\kappa(\omega^\prime)}{2\pi}} [\hat{a}_{\ell} \hat{A}_{\ell }^\dagger(\omega^\prime) + {\rm h.c.} ].
	\end{equation}
		The auxiliary field modes $\hat{A}_{\ell}$ and $\hat{B}_{\ell}$ with $\ell \!\! =\!\! 1, \dots, 4$ individually satisfy the usual bosonic commutation relations  $\big[A_{\ell}(\omega^\prime), A_{\ell}^\dagger(\omega^{\dprime})\big] \!\!=\!\! \big[B_{\ell}(\omega^\prime ), B_{\ell}^\dagger(\omega^{\dprime})\big] \!\! =\!\! \delta(\omega^\prime  - \omega^{\dprime} )$. For each optical mode, $\hat{\mathcal{O}}_{\ell, \omega^\prime} \in \{\hat{A}_{\ell,\omega^\prime},\hat{B}_{\ell.\omega^\prime}\}$, we define its effective input and output field as
	\begin{align}
		\hat{\mathcal{O}}_{\ell, \mrm{in}}(t) &:= - \frac{\i}{\sqrt{2\pi}} \int_{-\infty}^{\infty} \!\d\omega^{\prime}\, \hat{\mathcal{O}}_{\ell, \omega^\prime}(t_i)\, e^{-\i \omega^\prime (t - t_i)},\nonumber \\
		\hat{\mathcal{O}}_{\ell, \mrm{out}}(t) &:= - \frac{\i}{\sqrt{2\pi}} \int_{-\infty}^{\infty} \! \d\omega^{\prime}\, \hat{\mathcal{O}}_{\ell, \omega^\prime}(t_f)\, e^{-\i \omega^\prime (t - t_f)},   \label{eq:InputOutputField}
	\end{align}
	which satisfy $\left[\hat{\mathcal{O}}_{\ell, \mrm{in/out}}(t),\hat{\mathcal{O}}_{\ell, \mrm{in/out}}(t')^\dagger\right]=\delta(t-t')$ and lead to the following  input-output relation \cite{Steck}
	\begin{equation}\label{eq:IOrelation}
		\hat{\mathcal{O}}_{\ell, \mrm{out}}(t)  =  \hat{\mathcal{O}}_{\ell, \mrm{in}}(t) - \sqrt{\kappa} \hat{a}_{\ell}(t).
	\end{equation}

	Assuming all the cavity modes to be of same frequency $\omega_{\ell} \!\!=:\!\! \omega_{0}$ $\forall$ $\ell\!\!=\!\!1,\dots,4$, the cavity mode vector $\qvec{a} \!\!:=\!\! ( \hat{a}_{1}, ~\hat{a}_{2}, ~ \hat{a}_{3}, ~\hat{a}_{4} )^\T$ obeys the following Langevin equation~\cite{Perina2022Quantum,Perina2023Unavoidability,Naikoo2023,Wakefield2024,Perina2024Multiple,Thapliyal2024Multiple}:

\begin{equation}\label{eq:EOM_Time}
		\partial_t \qvec{a} = -\i (\omega_{0} \id + \mat{H})~ \qvec{a} + \mat{K}_\mrm{probe} \qvec{A}_\mrm{in} + \mat{K}_\mrm{diss} \qvec{B}_\mrm{in} + \mat{K}_\mrm{amp} \qvec{B}_\mrm{in}^\dagger,
\end{equation}
where 
\begin{equation}\label{eq:H4x4}
								\mat{H}   = \begin{pmatrix}
									-\i \gamma_1  &       g       	  &   J   			&   0\\
									g       &     \i \gamma_2   & 0    			&  J       \\
									J     &       0     		&  -\i \gamma_3   &  g  \\
									0     &        J	&   g   		   &     \i \gamma_4 
								\end{pmatrix}\!,
\end{equation}
and $\gamma_1 = (\eta_1 +  \kappa)/2 $, $\gamma_2 = (\eta_2 -  \kappa)/2 $, $\gamma_3 = (\eta_3 +  \kappa)/2 $ and $\gamma_4 = (\eta_4 -  \kappa)/2$. The input vectors as defined in \eqnref{eq:InputOutputField} are $\qvec{A}_\mrm{in} = ( \hat{A}_\mrm{1, in} ~  \hat{A}_\mrm{2, in}~  \hat{A}_\mrm{3, in}~  \hat{A}_\mrm{4, in} )^\T$, $\qvec{B}_\mrm{in} = ( \hat{B}_\mrm{1, in} ~  \hat{B}_\mrm{2, in}~  \hat{B}_\mrm{3, in}~  \hat{B}_\mrm{4, in} )^\T$, $\qvec{B}_\mrm{in}^{\dagger} = ( \hat{B}_\mrm{1, in}^\dagger ~  \hat{B}_\mrm{2, in}^\dagger~  \hat{B}_\mrm{3, in}^\dagger~  \hat{B}_\mrm{4, in}^\dagger )^\T \ne (\qvec{B}_\mrm{in})^\dagger$. The various coupling matrices are $	\mat{K}_\mrm{probe} = \mrm{diag} (\sqrt{\kappa}, \sqrt{\kappa}, \sqrt{\kappa}, \sqrt{\kappa})$, $\mat{K}_\mrm{diss} =  \mrm{diag} (\sqrt{\eta_1},0, \sqrt{\eta_3} , 0)$, $\mat{K}_\mrm{amp} =  \mrm{diag} (0, -\sqrt{\eta_2},0, -\sqrt{\eta_4})$.\bigskip

Up to this point, we have assumed that all system parameters are fixed and known. However, in realistic scenarios, certain parameters might be subject to small variations or uncertainties. To capture this, we now introduce a \emph{linear} perturbation parameter $\theta$, which is encoded within the system’s dynamical generator \(\mat{H}\) given in \eqnref{eq:H4x4}. The resulting perturbed generator, denoted by  \(\mat{H}_{\theta}\), takes the following form
\begin{equation}\label{eq:Htheta}
	\mat{H}_{\theta} = \mat{H} - \theta \, \mat{n}.
\end{equation}
Here, \(\mat{n}\) is the perturbation matrix that specifies which parameters in the system are affected by the perturbation. This matrix allows us to control the sensitivity of the system dynamics to changes in \(\theta\). 
		
		For instance, if we set \(\mat{n} = \id\), where \(\id\) is the identity matrix, then the perturbation parameter \(\theta\) influences the common cavity frequency, \(\omega_{0}\). This configuration implies that \(\theta\) shifts the entire frequency spectrum by a small amount, modeling a uniform frequency perturbation across all modes. In more general cases, \(\mat{n}\) could represent specific structures, allowing targeted perturbations to certain elements of \(\mat{H}\), such as coupling strengths, decay rates, or detuning parameters.
		
		Through this approach, we can analyze the robustness of the system's behavior under small parameter shifts and explore how different choices of \(\mat{n}\) impact the dynamics induced by the perturbed generator \(\mat{H}_{\theta}\).

\subsection{Solution in the Fourier space}
At this point, it is useful to transform \eqnref{eq:EOM_Time} into the  Fourier space  by defining the Fourier transform of a time-dependent operator $\qvec{u}(t) \!\!=\!\! (\hat{u}_{1}(t)~ \hat{u}_{2}(t) \dots )^\T$ as $\qvec{u}[\omega] \!\!:=\!\! \mathcal{F}_{\omega}[\qvec{u}(t)] \!\!=\!\! \int dt e^{\i \omega t} \qvec{u}(t)$. The correspondint position and mementum quadratures are defined as $\qvec{q}^{u}[\omega] \!\!=\!\! \qvec{u}[\omega] + (\qvec{u}[\omega])^\dagger$ and $i \qvec{p}^{u}[\omega] \!\!=\!\! \qvec{u}[\omega] - (\qvec{u}[\omega])^\dagger$. Thus, \eqnref{eq:EOM_Time} in the Fourier space reads:
\begin{align}
	- \i \omega \qvec{a}[\omega] = &-\i (\omega_{0} \id + \mat{H})~ \qvec{a}[\omega] + \mat{K}_\mrm{probe} \qvec{A}_\mrm{in}[\omega] \nonumber \\ &+ \mat{K}_\mrm{diss} \qvec{B}_{in}[\omega] + \mat{K}_\mrm{amp} \qvec{B}_\mrm{in}^\dagger[\omega]. \label{eq:EOM_Fourier}
\end{align}
The Fourier transform corresponds to a fixed unitary rotation in phase space~\cite{Simon2000}, under which the structure of the dynamical equations is preserved. As such, the generator $\smat{H}$ continues to act from the left on the operator vector $\hat{\qvec{a}}[\omega]$ in the frequency domain, just as in the time domain \cite{Ma1990}.  Equation~\eqref{eq:EOM_Fourier} therefore represents a change of basis rather than a change in the underlying dynamics.

Since we will be working with Gaussian probe fields, which are completely characterized by amplitude and covariance matrix, we define the quadratures vectors corresponding to the cavity modes $\qsvec{S}^{S}[\omega]  \!\!:=\!\! (\hat{q}_1 [\omega], \cdots, \hat{q}_4[\omega],  \hat{p}_1 [\omega], \cdots,  \hat{p}_4[\omega])^\T$, as well as the auxiliary field modes  $\qsvec{S}_\mrm{in}^{\bullet}[\omega]\!\!:=\!\!( \hat{q}^{\bullet_\mrm{in}}_1[\omega], \cdots,  \hat{q}^{\bullet_\mrm{in}}_4[\omega], \hat{p}^{\bullet_\mrm{in}}_1[\omega], \cdots,  \hat{p}^{\bullet_\mrm{in}}_4[\omega])^\T$, and $\bullet\!\!=\!\! A,B$.  In this quadrature representation \eqnref{eq:EOM_Fourier} becomes
\begin{align}
	-\omega\sform \qsvec{S}^{S}[\omega]  = &-(\omega_{0} \Id + \smat{H} ) \sform \qsvec{S}^{S}[\omega] + \sqrt{\kappa} \qsvec{S}^{A}_\mrm{in}[\omega] 
	\nonumber \\&+  \smat{K}_{B_1}\qsvec{S}^{B}_\mrm{in}[\omega] + \smat{K}_{B_2}\qsvec{S}^{B}_\mrm{in}[-\omega].
	\label{eq:dyn_phasespace}
\end{align}

	In the context of \eqnref{eq:EOM_Fourier}, any ($4\times4$) matrix $\mat{F}$ is now  expressed in the (symplectic) phase space as
\begin{equation}
	\smat{F}
	\equiv \stransf{\mat{F}}
	:= \begin{pmatrix}
		\Re[\mat{F}] & -\Im[\mat{F}] \\
		\Im[\mat{F}] & \Re[\mat{F}]
	\end{pmatrix},
	\label{eq:phase_space_rep}
\end{equation}
so that $\smat{H} = \stransf{\mat{H}}$, $\Id = \stransf{\id}$ corresponds to a $8\times 8$ identity matrix, and $\sform\!\! := \!\! \stransf{\i\id}$ is the symplectic form, which plays the role of the imaginary unit. Similarly, the  input-output relation in  \eqnref{eq:IOrelation} for the probe field amplitude is given in quadrature basis by 
\begin{equation}
	\qsvec{S}_\mrm{out}^{A}[\omega] = \qsvec{S}_\mrm{in}^{A}[\omega ] - \sqrt{\kappa} \qsvec{S}_\mrm{in}^{S}[\omega].
	\label{eq:in-out_phasespace}
\end{equation}
To express the dynamics entirely in terms of the input and output fields, we eliminate the internal system operator $\qsvec{S}^{S}$ by substituting the input-output relation~\eqnref{eq:in-out_phasespace} into the system's dynamical equation~\eqnref{eq:dyn_phasespace}. This yields a linear relation of the form $((\omega - \omega_{0}) \Id - \smat{H})\smat{S}_\mathrm{out}^{A}[\omega] = \cdots$, whose solution naturally involves the inverse $((\omega - \omega_{0}) \Id - \smat{H})^{-1}$. This inverse acts as the Green’s function of the system and encapsulates its frequency-dependent response to external inputs.

Substituting~\eqnref{eq:in-out_phasespace} into~\eqnref{eq:dyn_phasespace}, we obtain
\begin{align}\label{eq:SAout}
	\qsvec{S}_\mrm{out}^{A}[\omega] = &(\Id  - \smat{K}_\mrm{probe} \smat{G}[\omega]) \qsvec{S}_\mrm{in}^{A}[\omega] \nonumber \\&- \smat{K}_\mrm{probe} \smat{G}[\omega] \left( \smat{K}_\mrm{diss} \qsvec{S}_\mrm{in}^{B}[\omega] + \smat{K}_\mrm{amp}  \qsvec{S}^{B}_\mrm{in}[-\omega] \right),
\end{align}
where the coupling matrices are given by $\smat{K}_\mrm{probe} = \mrm{diag} (\mat{K}_\mrm{probe}, \mat{K}_\mrm{probe} )$, $\smat{K}_\mrm{diss} = \mrm{diag} ( \mat{K}_\mrm{diss},  \mat{K}_\mrm{diss})$ and $\smat{K}_\mrm{amp} = \mrm{diag} ( \mat{K}_\mrm{amp},  -\mat{K}_\mrm{amp})$, and $\smat{G}[\omega]$ is the system's response  function 
\begin{equation}
	\smat{G}[\omega] = \sform \bigg((\omega-\omega_0)\Id - \smat{H}\bigg)^{-1}.
	\label{eq:Greens}
\end{equation}
We can further  simplify \eqnref{eq:SAout} as  
\begin{align}
	\qsvec{S}_\mrm{out}^{A}[\omega] &= (\Id  - \smat{K}_\mrm{probe} \smat{G}[\omega]) \qsvec{S}_\mrm{in}^{A}[\omega]  - \smat{K}_\mrm{probe} \smat{G}[\omega] \smat{\mathcal{K}} \qsvec{S}^{B}[\pm \omega], \label{eq:SAoutSimplified}
\end{align}
where $\mathcal{K}$ is an $8 \times 16$ matrix given in block form by $\mathcal{K} \!\!=\!\! [ \mat{K}_\mrm{diss}, \mat{0}, \mat{K}_\mrm{amp}, \mat{0}; \mat{0},  \mat{K}_\mrm{diss},  \mat{0},  \mat{K}_\mrm{amp}]$ and the vector   $\qsvec{S}^{B}[\pm \omega] \!\!=\!\!  \big(\hat{q}_1^{B}[\omega] \cdots  \hat{q}_4^{B}[\omega],  \hat{q}_1^{B}[-\omega] \cdots \hat{q}_4^{B}[-\omega],\\  \hat{p}_1^{B}[\omega] \cdots \hat{p}_4^{B}[\omega], \hat{p}_1^{B}[-\omega] \cdots \hat{p}_4^{B}[-\omega]\big)^\T$ include both positive and negative frequency components.

We define the mean value as $\svec{S} \!:=\! \langle \qsvec{S} \rangle$, and assume that the auxiliary modes $\hat{B}_{\ell}$ are initialized in the vacuum state. Under this assumption, the contribution from $\svec{S}_\mathrm{in}^{B}[\pm \omega]$ vanishes in \eqnref{eq:SAoutSimplified}, leading to
\begin{equation}\label{eq:Sout}
     \svec{S}_\mrm{out}^{A}[\omega] = (\Id  - \smat{K}_\mrm{probe} \smat{G}[\omega]) \svec{S}_\mrm{in}^{A}[\omega].
\end{equation}
The covariance matrix of the output modes being measured, i.e.~$\smat{V}_\mrm{out}^A[\omega]$  has elements $[\smat{V}_\mrm{out}^A]_{jk}=\frac{1}{2} \qmean{ \{[\qsvec{S}_\mrm{out}^{A}]_j, [\qsvec{S}_\mrm{out}^{A}]_k\}}  - \qmean{[\qsvec{S}_\mrm{out}^{A}]_j}\qmean{[\qsvec{S}_\mrm{out}^{A}]_k}$. It is not hard to show that an input-output relation, similar to \eqnref{eq:Sout}, is given by \cite{Naikoo2023}
\begin{align}\label{eq:Vout}
	\smat{V}_\mrm{out}^{A}[\omega] &= \left(\Id -  \smat{K}_\mrm{probe} \smat{G}[\omega]\right) \smat{V}_\mrm{in}^{A}[\omega] \left(\Id -  \smat{K}_\mrm{probe} \smat{G}[\omega]\right)^\T \nonumber \\&+  \smat{K}_\mrm{probe} \smat{G}[\omega] \mathcal{K} \tilde{\smat{V}}_\mrm{in}^{B}[\omega] \mathcal{K}^\T \smat{G}[\omega]^\T.
\end{align}
In what follows, we use the input covariance matrices 
	\begin{equation}
		\smat{V}_\mrm{in}^{A} =  (2 n_\mrm{A} + 1) \Id_{8\times 8},\quad  \tilde{\smat{V}}_\mrm{in}^{B} =  (2n_\mrm{B} + 1) \Id_{16 \times 16},  \label{eq:InputCovMatrix}
	\end{equation}
	where $n_{A}$ and $n_{B}$ are the average thermal photons associated with the probe and bath modes, respectively.

The perturbation described in \eqnref{eq:Htheta}  translates in the Fourier space to the following
\begin{equation}
\smat{H}_{\theta} = \smat{H} - \theta \, \smat{n},
\end{equation}
where $\smat{n}$ is the matrix $\mat{n}$ in the Fourier space given by \eqnref{eq:phase_space_rep}. 
\noindent
	The role of $\mat{n}$ in shaping the structure and target of the perturbation was discussed earlier; in Fourier space, $\smat{n}$ serves the analogous purpose. It determines which spectral components or mode couplings are affected by the parameter $\theta$. For example, if $\mat{n} = \id$ introduces a uniform frequency shift in real space, the corresponding $\smat{n}$ yields a uniform spectral shift across all modes. More generally, $\smat{n}$ can be designed to encode structured perturbations in the spectral domain, allowing selective modification of specific modes or interactions.  Consequently, the $\theta$-dependent response function will be denoted by  $\smat{G}_{\theta} [\omega] = \sform \big((\omega-\omega_0)\Id + \theta \smat{n}  - \smat{H}\big)^{-1}$. If we further assume that the probe frequency is in resonance with the system frequency i.e. $\omega = \omega_{0}$, we have 
\begin{equation}
	\smat{G}_{\theta}[\omega = \omega_{0}] = \sform \big( \theta \smat{n}  - \smat{H}\big)^{-1}.
	\label{eq:GreensPerturbed} 
\end{equation}
As a result, the output amplitude in \eqnref{eq:Sout} and the covariance matrix in \eqnref{eq:Vout} depend on $\theta$ and we denote them as 
\begin{equation}\label{eq:SthetaVtheta}
	  \svec{S}_\mrm{out}^{A}[\omega = \omega_{0}] \rightarrow  \svec{S}_\mrm{out, \theta}^{A} , \quad   \svec{V}_\mrm{out}^{A}[\omega = \omega_{0}] \rightarrow \svec{V}_\mrm{out, \theta}^{A}.
\end{equation}
 We stress that the $\theta$-parameter dependence of $ \svec{S}_\mrm{out, \theta}^{A}$ and  $\svec{V}_\mrm{out, \theta}^{A}$ is entirely through the response function $\smat{G}_{\theta}[\omega = \omega_{0}]$ in \eqnref{eq:GreensPerturbed}.

	\subsection{Gaussian estimation of linear perturbations}\label{sec:GaussianEstimation}
	Since we are focusing on Gaussian probe signals -- the input bosonic modes $\hat{A}_{k,in}$ in \figref{fig:model4modes}, it is worth to briefly revisit  the estimation theory with  Gaussian states and Gaussian measurements \cite{Weedbrook2012,Ferraro2005}. A Gaussian state $\rho(\smat{S}, \smat{V})$ is completely characterized by its amplitude $\smat{S}$ and covariance matrix $\smat{V}$.  In the Gaussian estimation theory, an unkown parameter $\theta$ is encoded on these amplitude and covariance matrix resulting in a Gaussian state  $\rho_{\theta} = \rho(\smat{S}_{\theta}, \smat{V}_{\theta})$.  The ultimate precision in estimating $\theta$ is then given by  the QFI
	\begin{equation}\label{eq:defGaussiaQFI}
		\mathcal{F}_{\theta} = \frac{1}{2}\Tr\left[  \frac{d\smat{V}_{\theta}  }{d \theta} \smat{V}_{\theta}^{-1}  \frac{d \smat{V}_{\theta}  }{d \theta} \smat{V}_{\theta}^{-1}  \right] + 2 \left(\frac{d \smat{S}_{\theta}}{d \theta}\right)^T \smat{V}_{\theta}^{-1} \left(\frac{d \smat{S}_{\theta}}{d \theta}\right).
	\end{equation}
	The expression in Eq.~\eqref{eq:defGaussiaQFI} is the standard formula for the QFI in single-parameter estimation with Gaussian states, where the parameter $\theta$ is encoded in both the displacement vector $\smat{S}_{\theta}$ and the covariance matrix $\smat{V}_{\theta}$. A rigorous derivation of this result can be found in Ref.~\cite{Zhang2014}, to which we refer the interested reader. A generalization to the multiparameter case—though not required in the present work—has been developed in Ref.~\cite{Naikoo2023}.

	In the context of our model with the amplitude and covariance matrix given in \eqnref{eq:SthetaVtheta}, the QFI, as given in \eqnref{eq:defGaussiaQFI}, reads
	\begin{align}\label{eq:QFImodel}
		\mathcal{F}_{\theta} = &\frac{1}{2}\Tr\left[  \frac{d\smat{V}_\mrm{out,\theta}^{A}  }{d \theta} (\smat{V}_\mrm{out,\theta}^{A})^{-1}  \frac{d \smat{V}_\mrm{out,\theta}^{A}  }{d \theta} (\smat{V}_\mrm{out,\theta}^{A})^{-1}  \right] \nonumber  \\&+ 2 \left(\frac{d \smat{S}_\mrm{out,\theta}^{A}}{d \theta}\right)^T (\smat{V}_\mrm{out,\theta}^{A})^{-1} \left(\frac{d \smat{S}_\mrm{out,\theta}^{A}}{d \theta}\right).
	\end{align}
	One can use this formula directly to determine precision by fixing system parameters in $\smat{H}$ and  choosing a desired perturbation matrix $\smat{n}$. It is worth mentioning that the QFI is an optimized quantity which is obtained by maximizing the classical Fisher information over all possible measurements \cite{BraunsteinCaves1994}.  Thus, it is natural to ask if a specific measurement strategy saturates the precision obtained by QFI. As it is described in the next section, the  \emph{heterodyne} detection performed on the output signals is the optimal measurement for our system. Such a detection leads to a probability distribution $p(\smat{x}_{\theta})$ where $\smat{x}_{\theta}$ is a vector of (real) measurement outcomes. In the context of our bosonic system, the distribution reads
	\begin{equation}
		p(\smat{x}_{\theta}) = \frac{1}{(2 \pi)^{N} \sqrt{\smat{\smat{C}_{\theta}}}} \exp\left( - \frac{1}{2}(\smat{x}_{\theta} - \bar{\smat{x}})^\T \smat{C}_{\theta}^{-1}  (\smat{x}_{\theta} - \bar{\smat{x}}) \right),
	\end{equation}
	where
	\begin{equation}
	\smat{x}_{\theta} = \smat{S}_\mrm{out, \theta}^\mrm{A}, \quad 	\smat{C}_{\theta} = \smat{V}_\mrm{out, \theta}^{A} + \Id, \label{eq:HeteroCovMatrix}
	\end{equation}
	 is the amplitude and covariance matrix and the Fisher information is accordingly given by \cite{KayBook}
		\begin{align}\label{eq:CFImodel}
		F_{\theta} = &\frac{1}{2}\Tr\left[  \frac{d \smat{C}_{\theta}}{d \theta} \smat{C}_{\theta}^{-1}  \frac{d \smat{C}_{\theta}}{d \theta} \smat{C}_{\theta}^{-1}  \right] + 2 \left(\frac{d \smat{x}_{\theta}}{d \theta}\right)^T \smat{C}_{\theta}^{-1} \left(\frac{d \smat{x}_{\theta} }{d \theta}\right).
	\end{align}
	 \begin{figure}[h!]
		\includegraphics[width=\linewidth]{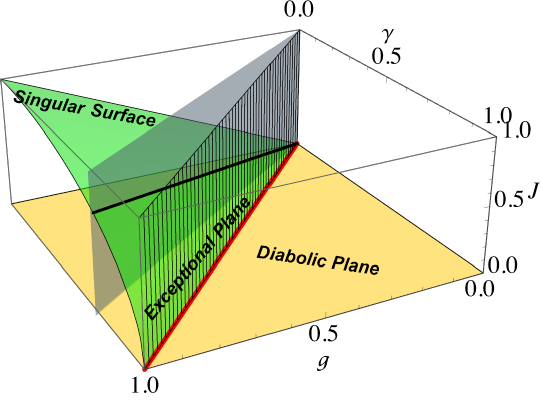}
		\caption{\textbf{Parameter space of the dynamical generator $\mat{H}$ in \eqnref{eq:H4x4}} with balanced gain and loss $\gamma_{\ell} \!\!=\!\! \gamma$ for all $\ell\!\! = \!\! 1,\dots,4$. Here, $J$ represents the coupling between successive modes with gain or loss, and $g$ is the coupling strength between a loss mode $\hat{a}_{1(3)}$ and a gain mode $\hat{a}_{2(4)}$. The green surface, denoted as \emph{singular surface}, shows points where $J = \sqrt{g^2 - \gamma^2}$, where $\mat{H}$ becomes \emph{singular} (denoted by $\mat{H}_\mathrm{S}$) as given in \eqnref{eq:Hsingular}. The yellow horizontal plane $J = 0$ contains \emph{diabolic points}, while the vertical meshed plane $g = \gamma$ includes \emph{exceptional points}. The intersection of singular surface, the exceptional plane and the diabolic plane is given by the hybrid exceptional-diabolic (HED) line in red. The gray plane at $g = \sqrt{2} \gamma$ is an arbitrary non-EP plane that intersects the singularity surface along the black line. The singular points along red (HEP) and black (non-HEP or non-EP) lines exhibit distinct scaling types of behavior as shown in \figref{fig:error} by red and black curves, respectively.		 		
		}
		\label{fig:singularity}
	\end{figure}

	\subsection{Precision limits from generator singularity}

	So far, we have laid the groundwork by showing that the precision (or QFI) can be computed through a straightforward formula \eqref{eq:QFImodel} involving the amplitude and covariance matrix. We are now prepared to explore the effect of the nature of the dynamical generator, denoted by $\smat{H}$ (or equivalently $\mat{H}$), on this precision. The precision is influenced by $\smat{H}$ via the response function  in \eqnref{eq:GreensPerturbed}, which serves as the central object of our analysis. Two distinct cases emerge:
	
	\emph{Non-singular dynamical generator.---} When $\smat{H}$ in \eqnref{eq:GreensPerturbed} [or equivalently $\mat{H}$ in \eqnref{eq:H4x4}] is invertible, the Neumann series can be applied to obtain
	\begin{equation}
		\smat{G}_{\theta}[\omega = \omega_{0}] = \sform \left( \theta \smat{H}_{\theta} - \smat{H} \right)^{-1} = - \sform \smat{H}^{-1} \sum\limits_{k=0}^{\infty} (\theta \smat{n}_{\theta} \smat{H}^{-1})^{k},
	\end{equation}
	which approaches ($- \sform \smat{H}^{-1}$) as $\theta \to 0$. Using this in \eqnref{eq:QFImodel},  the QFI  turns out to be
	\begin{equation}
		\mathcal{F}_{\theta} = a_{0} + \order{\theta}, \quad a_0 \ne 0,
	\end{equation}
	indicating that the leading-order term is independent of $\theta$. Here, we have
	\begin{equation}
		a_{0} = \frac{1}{2} \Tr \left[ \smat{Y}^{-1} \smat{Z} \smat{Y}^{-1} \smat{Z} \right] + 2 \kappa^2 \smat{S}_\mathrm{in}^{\T} g_{1}^{\T} \smat{Y}^{-1} g_{1} \smat{S}_\mathrm{in},
	\end{equation}
	where
	\begin{align}
		\smat{Y} &= \smat{V}_\mathrm{in}^{A} - \kappa (g_{0} \smat{V}_\mathrm{in}^{A} + \smat{V}_\mathrm{in}^{A} g_{0}^{\T}) + g_{0} \Lambda g_{0}^{\T},\\
		\smat{Z} &= -\kappa (g_1 \smat{V}_\mathrm{in}^{A} + \smat{V}_\mathrm{in}^{A} g_1^{\T}) + g_{0} \Lambda g_{1}^{\T} + g_1 \Lambda g_{0}^{\T}, \\
		\Lambda &= \kappa^2 \smat{V}_\mathrm{in}^{A} + \kappa \mathcal{K} (\smat{V}_\mrm{in}^\mrm{B_1} \oplus \smat{V}_\mrm{in}^\mrm{B_2}) \mathcal{K}^{\T},
	\end{align}  
with $g_k = - \smat{J} \smat{H} (\smat{n} \smat{H}^{-1})^{k}$ for $k=0,1$~\cite{Naikoo2023}. The auxiliary matrices $\smat{Y}$, $\smat{Z}$, and $\Lambda$ are introduced to express the QFI prefactor $a_0$ in a compact and analytically manageable form. These matrices capture the combined effects of the input probe noise via $\smat{V}_\mrm{in}^{A}$, and the environmental noise contributions, characterized by $\smat{V}_\mrm{in}^{B_1}$ and $\smat{V}_\mrm{in}^{B_2}$. This clarifies how the different noise sources influence the QFI.

	\emph{Singular dynamical generator.---} Conversely, when $\smat{H}$ is non-invertible, or \emph{singular}, the inverse of \eqnref{eq:GreensPerturbed} can be calculated using the Sain-Massey expansion from singular matrix perturbation theory \cite{Sain1969,FeiZhou}, yielding
	\begin{align}\label{eq:SMexpansion}
		\smat{G}_{\theta}[\omega = \omega_{0}] &= \sform \left( \theta \smat{n} - \smat{H}_\mathrm{S} \right)^{-1} \nonumber \\
		&= \theta^{-s} \left[\X_{0} + \theta \X_{1} + \theta^2 \X_{2} + \cdots \right],
	\end{align}
	where $\X_{0} \ne 0$, and $s$ is an integer that determines the pole order in this expansion (see Appendix \ref{app:SM}). In this scenario, the QFI in \eqnref{eq:defGaussiaQFI} exhibits a divergent behavior,
	\begin{equation}
		\mathcal{F}_{\theta} = \theta^{-2s} \left[ b_{0} + \order{\theta} \right], \quad b_{0} \ne 0, \label{eq:QFIsingular}
	\end{equation}
	where the coefficient is given by \cite{Naikoo2023}
	\begin{align}\label{eq:b0}
		b_{0} &= \Tr \left( \smat{n} \X_{0} \smat{n} \X_{0} + \smat{V}_\mathrm{in}^{-1} \smat{n} \X_{0} \smat{V}_\mathrm{in} \X_{0}^{\T} \smat{n}^{\T} \right) \nonumber \\&+ \kappa^2 \svec{S}_\mathrm{in}^{\T} \X_{0}^{\T} \smat{n}^{\T} \smat{V}_\mathrm{in}^{-1} \smat{n} \X_{0} \svec{S}_\mathrm{in}.
	\end{align}
	The above analysis  reveals that the nature of the dynamical generator $\smat{H}$ critically affects the behavior of the QFI and, consequently, the achievable precision in parameter estimation.  In essence, the nature of the dynamical generator $\smat{H}$—whether singular or non-singular—determines the balance between stable precision and enhanced sensitivity in parameter estimation. For non-singular $\smat{H}$, the QFI remains finite as $\theta \to 0$, ensuring constant precision and resilience to small perturbations, which supports stable measurement outcomes. Conversely, when $\smat{H}$ is singular, the QFI diverges as $\theta \to 0$, allowing for high sensitivity to parameter variations. This distinction enables the design of quantum systems tailored for either stable precision or heightened sensitivity, aligning experimental goals with the nature of $\smat{H}$.

		\section{Enhanced Precesion at Hybrid Exceptional-Diabolic points}\label{sec:Analysis}
		Let us first take a quick look at the structure of the dynamical generator $\mat{H}$ in \eqnref{eq:H4x4}. If we assume all the gain and loss rates are equal i.e. $\gamma_{\ell} = \gamma$,  $\forall \ell = 1,2,3,4$,  the matrix $\mat{H}$ can be expressed as direct sum $	\mat{H} = \mat{M}_{1} \otimes \id + \id \otimes \mat{M}_{2}$ where $\mat{M}_{1} = [ 0, J; J, 0]$ and $\mat{M}_{2} = [- \i \gamma, g; g, \i \gamma]$ with eigenvalues $\lambda_{1, \pm} = \pm J$ and $\lambda_{2, \pm} = \pm \sqrt{g^2 - \gamma^2}$, respectively \cite{arkhipov2023dynamically}. Such a decomposition is useful since the eigenvalues of $\mat{H}$, denoted by $\mu_{k,\ell,\pm} = \lambda_{k, \pm} + \lambda_{\ell,\pm}$ with $k, \ell \!\!=\!\! 1,2$, are simply given by the sum of the eigenvalues of $\mat{M}_1$ and $\mat{M}_2$. Secondly, $\mat{H}$ demonstrates an exceptional curve when  $g = \gamma$  for all points  $\mu_{k,\ell,\pm} \!\!=\!\! \pm J$ while as for $J=0$ pertains to a diabolic curve given by all points $\mu_{k,\ell,\pm} \!\!=\!\! \pm \sqrt{g^2 - \gamma^2}$.  However, what  is going to be an even more interesting case for us is that $\mat{H}$ is singular for all points satisfying $J = \pm \sqrt{g^2 - \gamma^2}$ and we denote it as follows (restricting to positive square root)
		 \begin{equation}\label{eq:Hsingular}
		 	\mat{H}_\mrm{S}  = \begin{pmatrix} 
												 		-\i \gamma  & g & \sqrt{g^2-\gamma ^2} & 0 \\
												 			g & \i \gamma  & 0 & \sqrt{g^2-\gamma ^2} \\
												 			\sqrt{g^2-\gamma ^2} & 0 & -\i \gamma  & g \\
												 			0 & \sqrt{g^2-\gamma ^2} & g & \i \gamma  \\
												 	     \end{pmatrix}.
		 \end{equation}

		 \begin{figure}
			\includegraphics[width=\linewidth]{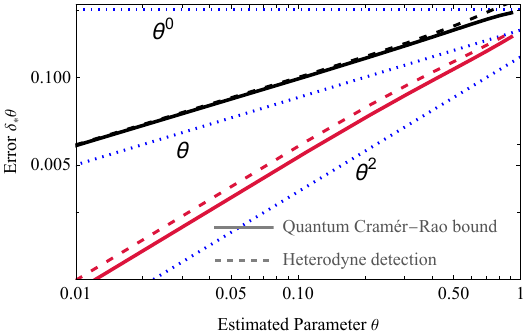}
			\caption{\textbf{Estimation error of the parameter $\theta$} affecting cavity frequency $\omega_0$ in \eqnref{eq:EOM_Time}. The solid curves correspond to the quantum Cramér-Rao bound  $\delta_{Q} \theta = 1/\sqrt{\mathcal{F}_{\theta}}$, where $\mathcal{F}_{\theta}$ is numerically calculated from  \eqnref{eq:QFImodel}. The dashed curves represent the classical Cramér-Rao bound achieved through heterodyne detection, expressed as $ \delta_{C} \theta = 1/\sqrt{F_{\theta}}$ with $F_{\theta}$ obtained numerically from \eqnref{eq:CFImodel}. The scaling depicted by the quantum Cramér-Rao bound  in the two cases is in agreement with the analytic results given in~Eqs. (\ref{eq:CRBquad}) and (\ref{eq:CRBlinear}). Similarly analytic predictions are obtained for heterodyne detection.  The dotted blue lines serve as reference lines, scaling proportionally to $\theta$ and $\theta^2$, as indicated. 
				The quadratic error scaling exhibited by the red solid and dashed curves is achieved by tuning the singular generator $\mat{H}_S$ in \eqnref{eq:Hsingular} to $g = \gamma$, corresponding to the HED line in the parameter space shown in \figref{fig:singularity}. In contrast, the black solid and dashed curves indicate linear scaling, where the singular generator $\mat{H}_S$ is set to $g = \sqrt{2} \gamma$, representing the non-HED black line in \figref{fig:singularity}.
				The parameters used include a common loss/gain rate $\gamma$ set to $\gamma = 1$, coupling strengths $\kappa = 1$, $\eta_1 = \eta_2 = \eta_3 = \eta_4 = 0.5$. The average number of thermal photons in the input Gaussian probe and the bath is set to $n_{A} = n_{B}= 1$.
			}
			\label{fig:error}
		\end{figure}	
		 
		 In light of the discussion in \secref{sec:GaussianEstimation}, one expects enhanced sensitivity in sensing linear perturbations when the system is operator at $\mat{H}_\mrm{S}$. However, note that in  \secref{sec:GaussianEstimation}, the conclusion about enhanced precision was based on the singularity $\smat{H}$ and not of $\mat{H}$. However, it turns out that $\det \smat{H} = |\det \mat{H}|^{2}$ (see Lemma 2 in the supplementary material of  \cite{Naikoo2023}). Therefore, in light of \eqnref{eq:phase_space_rep}, $\smat{H}_\mrm{S} = \stransf{\mat{H}_\mrm{S}}$ is also singular and the expansion \eqnref{eq:SMexpansion} applies.

		 The precision enhancement is quantified in terms of the pole order $s$ in the expansion \eqnref{eq:SMexpansion} -- which for a particular $\smat{H}_\mrm{S}$ is determined by the perturbation matrix $\smat{n}_{\theta}$.  Let us consider the as an example the perturbation matrix $\smat{n} = \Id$ -- an $8\times 8$ identity matrix. This matrix perturbs the common cavity frequency $\omega_{0}$ by the same amount $\theta$ as $(\omega_{0} - \theta)$.

 \emph{Sensing at an {\rm HED} singularity.---} Let us consider the case when $g=\gamma$ i.e. the system is at HED point. Following the procedure in \cite{Sain1969,Naikoo2023} we find that the response function in \eqnref{eq:SMexpansion} has the following form
		 \begin{equation}
		  	 \smat{G}_{\theta}[\omega = \omega_{0}] = \theta^{-2} (\X_{0} + \theta \X_{1}), \label{eq:Gm2}
		 \end{equation}
		 where as shown in \appref{eqApp:SMexp} the coefficients are given by
		 \begin{equation}
            \X_{0} = \smat{H}_\mrm{S}|_{g = \gamma}, \quad \X_{1} = \Id. \label{eq:CoeffsHED}
		 \end{equation}
		 We can now use the response function $ \smat{G}_{\theta}[\omega = \omega_{0}] $ given by  \eqnref{eq:Gm2} to calculate the corresponding amplitude  and covariance matrix given in   \eqnref{eq:Sout} and \eqnref{eq:Vout}, respectively.  Subsequently, we use these quantities to calculate the QFI given in  \eqnref{eq:QFImodel}  and  obtain
		 \begin{equation}
		 	\mathcal{F}_{\theta}^\mrm{HED} = \theta^{-4} \left[ b_{0}^{\prime} + \order{\theta}\right], \label{eq:QFIm4}
		 \end{equation}
		 		where $b_{0}^{\prime}$ is $b_{0}$ given in \eqnref{eq:b0} for this particular case i.e. with perturbation matrix $\smat{n} = \smat{I}$ and coefficient $\smat{X}_{0}$ and $\smat{X}_{1}$ given in \eqnref{eq:CoeffsHED}. The precision in \eqnref{eq:QFIm4}  leads to quadratic scaling of the error 
		 		  \begin{equation}
		 		  	\delta_{Q} \theta = \frac{1}{\sqrt{\mathcal{F}_{\theta}^\mrm{HED}} } \propto \theta^{2}. \label{eq:CRBquad}
		 		  \end{equation}

	\begin{table}[t]
		\centering
		\renewcommand{\arraystretch}{1.5} 
		\setlength{\tabcolsep}{4pt}      
		\begin{tabular}{ccc} 
			\textbf{Singularity Region} & \textbf{Condition} & \textbf{Sensitivity $(\delta_{Q}\theta, \delta_{C}\theta)$} \\ 
			\hline 
			\hline
			Singular Surface  & $J^2 = g^2 - \gamma^2$ & $\theta~(g \ne \gamma)$, $\theta^2~(g=\gamma)$ \\ 
			Diabolical Plane  & $J = 0$                & $\theta^{0}$ if $g \ne \gamma$                \\ 
			Exceptional Plane & $g = \gamma$           & $\theta^{0}$ if $J \ne 0$                     \\ 
			HED Line          & $(g, \gamma, J) = (\gamma, \gamma, 0)$ & $\theta^2$ \\ 
			\hline 
		\end{tabular}
		\caption{ \textbf{Summary of singularity types}, their corresponding conditions, and their relevance to the sensing parameter $\theta$, which influences the common mode frequency. The sensitivity is quantified by the quantum and classical Cramér-Rao bounds, represented as $\delta_{Q} \theta = 1/\mathcal{F}_{\theta}$ and $\delta_{C} \theta = 1/F_{\theta}$, where $\mathcal{F}_{\theta}$ and $F_{\theta}$ denote the quantum and classical (pertaining to heterodyne detection) Fisher information, as provided in Eqs.~(\ref{eq:QFImodel}) and (\ref{eq:CFImodel}), respectively.}
		\label{tab:singularity_types}
	\end{table}

 \emph{Sensing at a non-{\rm HEP} singularity.---}  Next, let us consider a non-HEP singularity by setting $g = \sqrt{2} \gamma$ in \eqnref{eq:Hsingular} [or equivalently in its quadrature representation $\smat{H}_{S}(g,\gamma)$], we find
 \begin{equation}
 	\smat{G}_{\theta}[\omega = \omega_{0}] = \theta^{-1} \left[\X_{0} + \theta \X_{1} + \theta^{2} \X_{2} + \cdots \right],
 \end{equation}
 where the even  and odd  coefficients are given by $\X_{2k} = (2\gamma)^{-2k} \X_\mrm{even}$ and $\X_{2k+1} =  (2\gamma)^{-2k-1} \X_\mrm{odd}$
 
 \begin{small}
 	\begin{align}
 		\X_\mrm{even} &= 
 		\begin{pmatrix}
 			r^2 & 0    & 0 & r & 0 & 0 & -r^2 & 0 \\
 			0    & r^2 & r & 0 & 0 & 0 & 0 & r^2  \\
 			0    & r     &  r^2 & 0 & - r^2 & 0 & 0 & 0 \\
 			r  & 0 & 0 & r^2 & 0 & r^2 & 0 & 0 \\
 			0 & 0 & r^2 & 0 & r^2 & 0 & 0 & r \\
 			0 & 0 & 0 & -r^2 & 0 & r^2 & r & 0 \\
 			r^2 & 0 & 0 & 0 & 0 & r & r^2 & 0 \\
 			0 & - r^2 & 0 & 0 & r & 0 & 0 & r^2 \\
 		\end{pmatrix}, \nonumber \\
 		\X_\mrm{odd} &= 
 		\begin{pmatrix}
 			0 & r & -r^2 & 0 & -r^2 & 0 & 0 & 0 \\
 			r & 0 & 0 & -r^2 & 0 & r^2 & 0 & 0 \\
 			-r^2 & 0 & 0 & r & 0 & 0 & -r^2 & 0 \\
 			0 & -r^2 & r & 0 & 0 & 0 & 0 & r^2 \\
 			r^2 & 0 & 0 & 0 & 0 & r & -r^2 & 0 \\
 			0 & -r^2 & 0 & 0 & r & 0 & 0 & -r^2 \\
 			0 & 0 & r^2 & 0 & - r^2 & 0 & 0 & r \\
 			0 & 0 & 0 & - r^2 & 0 & - r^2 & r & 0 \\
 		\end{pmatrix}.  \label{eq:Coeffs}
 	\end{align}
 \end{small}
 where $k=0,1,2,\dots$ and  $r=-1/\sqrt{2}$.  Thus, in this case, the precision behaves as 
 \begin{equation}
\mathcal{F}_{\theta}^\mrm{non\!-\!HED} = \theta^{-2} \left[b_{0}^{\dprime} + \order{\theta}\right], \label{eq:QFIm2}
 \end{equation} 
 with $b_{0}^{\dprime}$ being $b_{0}$ in \eqnref{eq:b0} with coefficients given in \eqnref{eq:Coeffs}. 
 The form of precision in \eqnref{eq:QFIm2} leads to a linear scaling of the error 
 \begin{equation}
 	\delta_{Q} \theta = \frac{1}{\sqrt{\mathcal{F}_{\theta}^\mrm{non\!-\!HED}}} \propto \theta. \label{eq:CRBlinear}
 \end{equation}
Consequently, while perturbing the system at the HED singularity achieves a quadratic scaling of error, perturbations away from this singularity result in only linear error scaling. These findings align with the numerical results shown in \figref{fig:error}, which were obtained by directly computing the QFI from \eqnref{eq:defGaussiaQFI} without any approximations.  For heterodyne detection, the CFI is given by \eqnref{eq:CFImodel} using the amplitude and covariance matrix specified in \eqnref{eq:HeteroCovMatrix}. By following the same procedure as outlined above, and analogous to Eqs.~(\ref{eq:QFIm4}) and (\ref{eq:QFIm2}), the CFI in the two cases is obtained as:
	\begin{equation}
		F_{\theta}^\mrm{HED}  = \theta^{-4} \left[ c_{0}^{\prime} + \order{\theta} \right], \quad F_{\theta}^\mrm{non\!-\!HED}  = \theta^{-2} \left[c_{0}^{\dprime} + \order{\theta}\right], \label{eq:CFIm2m4}
	\end{equation}
	where the coefficients $c_{0}^{\prime} > 0$ and $c_{0}^{\dprime} > 0$. Consequently, the corresponding errors scale as:
	\begin{equation}
		\delta_{C}\theta = \frac{1}{\sqrt{F_{\theta}^\mrm{HED}}} \propto \theta^{2}, \quad \delta_{C}\theta = \frac{1}{\sqrt{F_{\theta}^\mrm{non\!-\!HED}}} \propto \theta.   \label{eq:CRBquadLinear}
	\end{equation}
	The analytical predictions for the scaling behaviors in Eqs.~(\ref{eq:CRBquad}), (\ref{eq:CRBlinear}), and (\ref{eq:CRBquadLinear}) are supported by the plots in \figref{fig:error}, which were obtained by numerically computing the Fisher information given in Eqs.~(\ref{eq:QFImodel}) and (\ref{eq:CFImodel}). A summary of the scaling behavior in different situations is presented in~\tabref{tab:singularity_types}.

\subsection{Experimental feasibility} 
	The realization of the HED conditions $\gamma_\ell = \gamma$ and $J = \sqrt{g^2 - \gamma^2}$ relies on precise engineering of both coherent couplings and dissipative processes. Such control is increasingly achievable in photonic and circuit QED platforms. In particular, tunable gain and loss rates have been demonstrated using optical pumping~\cite{Peng2014ParityTime, Chang2014ParityTime} and reservoir engineering techniques~\cite{Metelmann2015,Mirhosseini2018}. Couplings between photonic modes or superconducting resonators can be dynamically tuned using integrated waveguide circuits or parametric modulation~\cite{Fitzpatrick2017,Blais2021}. These advances suggest that the parameter regimes required for observing HED-induced QFI enhancement are within reach of current experimental capabilities.

\subsection{Effect of measurement imperfections}
In practical implementations, various imperfections can influence the sensitivity enhancement predicted in our idealized model. First, our analysis already accounts for \emph{thermal noise} by considering a thermal input state, as defined in Eq.~(24), which serves as a realistic model for finite-temperature environments. However, other sources of technical noise may also play a role.

A key practical imperfection in quantum measurements is \emph{detector inefficiency}, which reduces the amount of information that can be extracted from a system. This effect can be captured in several equivalent ways. One common approach models inefficiencies using \emph{positive operator-valued measures} (POVMs), which generalize projective measurements and naturally describe imperfect detection processes~\cite{Wiseman2009Quantum, Holevo2011Probabilistic}. Alternatively, inefficiency can be modeled by passing the output quantum state through a fictitious \emph{lossy bosonic channel} before detection~\cite{Monras2007Optimal, Escher2011General, Ozdemir2001}, or equivalently, by inserting fictitious \emph{beam splitters} in front of ideal detectors, as discussed in~\cite{Bartkowiak2014}.

In the specific case of \emph{homodyne detection}, reduced detector efficiency manifests as a \emph{smoothing of the Wigner function}, due to the corresponding smoothing of its quadrature marginals relative to those obtained with ideal, unit-efficiency detectors~\cite{Leonhardt1993}. Across all these modeling frameworks, detector inefficiency results in a degradation of the QFI, which may even suppress its divergence near \emph{singular points} such as HED points~\cite{Zhou2018QEPQFI, Lau2018}.

Another relevant factor is \emph{gain-loss imbalance}, where the ideal symmetry $\gamma_\ell = \gamma$ may not be exactly satisfied. Such deviations can shift or even remove the system from a singular point, altering the pole structure of the response function and potentially reducing the achievable sensitivity. In extreme cases, an imbalance may destabilize the system, leading to runaway amplification or decaying modes.

At the modeling level, both thermal noise and gain-loss imbalance can be rigorously treated using general open-system frameworks, such as the Heisenberg-Langevin formalism (as we use here) or master equations with temperature-dependent Lindblad terms~\cite{Breuer2002TheTheory}. These approaches enable the description of nonzero-temperature baths and dissipative processes on equal footing, and can be used to systematically study stability and metrological performance.

While we focus here on analytic structure and ideal scaling, a more detailed numerical analysis of imperfections—including specific thermal models, detection inefficiencies, and HED robustness—could be fruitfully explored in a follow-up study aimed at guiding experimental realizations.

\section{Conclusion}\label{sec:Conclusion}
In conclusion, our findings highlight the persistence of quadratic scaling at second-order EPs, even in the presence of orthogonal subspace induced by the second-order DPs in a four mode bosonic system. The interplay between EP condition, the singularity condition, and the diabolic condition of the dynamical generator is critical to achieving this enhanced precision. Notably, the quadratic scaling is uniquely observed at HED points, where these conditions naturally coalesce. This underscores the significance of HED points in optimizing precision and offers valuable insights for advancing non-Hermitian systems and quantum sensing applications.  By using the tools of the singular matrix perturbation theory and Gaussian estimation framework, we reveal that HED points offer a significant twofold improvement in sensitivity over standard singularities in this system. The ultimate estimation limits, derived through the  quantum Fisher information, identify heterodyne detection as the optimal measurement strategy to achieve this enhanced scaling. These results deepen our understanding of HED singularities and open pathways for their application in high-precision quantum sensing and metrology.

\section*{Acknowledgement}
This work was supported by the Polish National Science Centre (NCN) under the Maestro Grant No. DEC-2019/34/A/ST2/00081. We thank Jan Ko\l{}ody\'{n}ski for insightful discussions.
	
\appendix
\setcounter{secnumdepth}{1} 

\section{Quantum Langevin equations}
The total Hamiltonian describing the sensor model in \figref{fig:model4modes} reads
\begin{equation}
	\hat{H} = \hat{H}_\mrm{S} + \hat{H}_\mrm{B} + \hat{H}_\mrm{SB},
\end{equation}
where
\begin{align}
	\hat{H}_\mrm{S} = &\sum\limits_{\ell = 1}^{4} \omega_k \hat{a}_{k}^{\dagger} \hat{a}_{k} + g (\hat{a}_{1}^{\dagger} \hat{a}_{2}  + \hat{a}_{3}^{\dagger} \hat{a}_{4}  + {\rm h.c.}) \nonumber \\&+ J (\hat{a}_{1}^{\dagger} \hat{a}_{3} + \hat{a}_{2}^{\dagger} \hat{a}_{4} + {\rm h.c.}),\\
	\hat{H} _\mrm{B} \approx & \sum\limits_{\ell = 1}^{4}   \int_{-\infty}^{\infty} d \omega^\prime~\omega^\prime ~  \hat{A}_{\ell }^\dagger( \omega^\prime) \hat{A}_{\ell}( \omega^\prime) \nonumber \\&+  \sum\limits_{\ell = 1}^{4}   \int_{-\infty}^{\infty} d \omega^\prime~\omega^\prime ~  \hat{B}_{\ell }^\dagger( \omega^\prime) \hat{B}_{\ell}( \omega^\prime), \\
	\hat{H}_\mrm{SB}  \approx  &\sum\limits_{\ell = 1}^{4}    \int_{-\infty}^{\infty} d \omega^\prime ~ \sqrt{\frac{\kappa(\omega^\prime)}{2\pi}} [\hat{a}_{\ell} \hat{A}_{\ell }^\dagger(\omega^\prime) + {\rm h.c.} ]  
	\nonumber \\&+ \sum\limits_{\ell = 2,4}  \int_{-\infty}^{\infty} d \omega^\prime ~ \sqrt{\frac{\eta_{\ell} (\omega^\prime)}{2\pi}} [  \hat{a}_{\ell}^{\dagger} \hat{B}_{\ell}^\dagger(\omega^\prime) + {\rm h.c.}]    
	\nonumber \\&+ \sum\limits_{\ell = 1,3}  \int_{-\infty}^{\infty} d \omega^\prime ~ \sqrt{\frac{\eta_{\ell} (\omega^\prime)}{2\pi}} [  \hat{a}_{\ell}  \hat{B}_{\ell}^\dagger(\omega^\prime)  + {\rm h.c.} ],
\end{align}
with commutation relations $\big[\hat{A}_{\ell}(\omega^\prime), \hat{A}_{\ell}^\dagger(\omega^{\dprime})\big] \!\!=\!\! \big[\hat{B}_{\ell}(\omega^\prime ), \hat{B}_{\ell}^\dagger(\omega^{\dprime})\big] \!\!=\!\! \delta(\omega^\prime  - \omega^{\dprime} )$. The Heisenberg equation of motion $\partial_t \hat{\mathscr{O} } \!\!=\!\! -i [\hat{\mathscr{O}}, \hat{H}_\mrm{total}]$, where $\hat{H}_\mrm{total} = \hat{H}_\mrm{S} + \hat{H}_\mrm{B} +  \hat{H}_\mrm{SB}$:
\begin{align}
	\partial_t \hat{a}_1 &= -i \omega_1 a_1 - i g_{12} a_2 - i J a_3 
	- \hat{f}_1^{A} - \hat{f}_1^{B}, \label{eq:a1dot} \\
	\partial_t \hat{a}_2 &= -i \omega_2 a_2 - i g_{12} a_1 - i J a_4 
	- \hat{f}_2^{A} - (\hat{f}_2^{B})^\dagger, \label{eq:a2dot}\\
	\partial_t \hat{a}_3 &= -i \omega_3 a_3 - i g_{34} a_4 - i J a_1 
	- \hat{f}_3^{A} - \hat{f}_3^{B}, \label{eq:a3dot}\\
	\partial_t \hat{a}_4 &= -i \omega_4 a_4 - i g_{34} a_3 - i J a_2 
	- \hat{f}_4^{A} - (\hat{f}_4^{B})^\dagger, \label{eq:a4dot}
\end{align}
with 
\begin{align}
	\hat{f}_j^{A} &= \frac{i}{\sqrt{2\pi}} \int_{-\infty}^{\infty} d\omega^\prime \sqrt{\kappa(\omega^\prime)} 
	\biggl[ \hat{A}_j(\omega^\prime)|_{t=t_0} e^{-i\omega^\prime (t-t_0)} \nonumber \\ 
	& \hspace{0.01cm} - i \sqrt{\frac{\kappa(\omega^\prime)}{2\pi}} \int_{t_0}^{t} dt^\prime a_j(t^\prime) e^{-i\omega^\prime (t-t^\prime)} \biggr], \\
	\hat{f}_j^{B} &= \frac{i}{\sqrt{2\pi}} \int_{-\infty}^{\infty} d\omega^\prime \sqrt{\eta_j(\omega^\prime)} 
	\biggl[ \hat{B}_j(\omega^\prime)|_{t=t_0} e^{-i\omega^\prime (t-t_0)} \nonumber \\ 
	& \hspace{0.01cm} - i \sqrt{\frac{\eta_j(\omega^\prime)}{2\pi}} \int_{t_0}^{t} dt^\prime a_j(t^\prime) e^{-i\omega^\prime (t-t^\prime)} \biggr].
\end{align}

We proceed by solving for the probe fields $\hat{A}_{k}$ and the reservoir fields $\hat{B}_{k}$ and substitute in Eqs. (\ref{eq:a1dot})-(\ref{eq:a4dot}). Subsequently, we invoke  the Markov approximation by ignoring the frequency dependence of the  rates
\begin{equation}
	\kappa (\omega^\prime) \approx \kappa, \quad \eta_k(\omega^\prime) \approx \eta_k.
\end{equation}
This leads to a simpler set of equations
\begin{align}
	\partial_t \hat{a}_1 &=   -i \omega_1 \hat{a}_1 - i g \hat{a}_2  - i J \hat{a}_3  + \hat{f}_{1}^{\prime}, \label{eq:a1dot2}\\
	\partial_t \hat{a}_2 &=    - i \omega_2 \hat{a}_2 - i g \hat{a}_1 - i J \hat{a}_4 + \hat{f}_{2}^{\prime},\label{eq:a2dot2}\\
	\partial_t \hat{a}_3 &=   -i \omega_3 \hat{a}_3 - i g \hat{a}_4  - i J \hat{a}_1  +  \hat{f}_{3}^{\prime}, \label{eq:a3dot2}\\
	\partial_t \hat{a}_4 &=   - i \omega_4 \hat{a}_4 - i g \hat{a}_3  - i J \hat{a}_2  + \hat{f}_{4}^{\prime}, \label{eq:a4dot2}
\end{align}
where
\begin{align}
	\hat{f}_{1}^{\prime} &=  \sqrt{\kappa} \hat{A}_\mrm{1,in}  - \frac{\kappa}{2} \hat{a}_1   +  \sqrt{\eta_1} \hat{B}_\mrm{1,in}  -  \frac{\eta_1}{2} \hat{a}_{1}, \label{eq:f1prime}\\
	\hat{f}_{2}^{\prime} &=  \sqrt{\kappa} \hat{A}_\mrm{2,in}  -  \frac{\kappa}{2} \hat{a}_{2}  - \sqrt{\eta_2} \hat{B}_\mrm{2,in}^\dagger   + \frac{\eta_2}{2} \hat{a}_2, \label{eq:f2prime}\\
	\hat{f}_{3}^{\prime} &= \sqrt{\kappa} \hat{A}_\mrm{3,in}  - \frac{\kappa}{2} \hat{a}_3  + \sqrt{\eta_3} \hat{B}_\mrm{3,in} - \frac{\eta_3}{2} \hat{a}_3,\label{eq:f3prime}\\
	\hat{f}_{4}^{\prime} &= \sqrt{\kappa} \hat{A}_\mrm{4,in}  -  \frac{\kappa}{2} \hat{a}_4  -  \sqrt{\eta_4} \hat{B}_\mrm{4,in}^\dagger  +  \frac{\eta_4 }{2} \hat{a}_4.\label{eq:f4prime}
\end{align}
The Eqs. (\ref{eq:a1dot2})-(\ref{eq:a4dot2}) together with Eqs. (\ref{eq:f1prime})-(\ref{eq:f4prime}) constitute the \eqnref{eq:EOM_Time}  in the main text written in a compact form with the definition of the input fields given in \eqnref{eq:InputOutputField} of the main text.

\section{The Sain-Massey procedure for expanding inverses of singular matrices}
\label{app:SM}
Let an $n \times n$ meromorphic matrix function $A(z)$ have a Laurent expansion about $z_0$:
\begin{equation}
	A(z) = \sum_{j=0}^{\infty} (z - z_0)^j A_j,
\end{equation}
where $A_j \in \mathbb{C}^{n \times n}$, $A_{0} \neq 0$. 
%
%
 Suppose the inverse $\A^{-1}(z)$ exists in some (possibly punctured) disc centered at $z = z_0$. Then, $\A^{-1}(z)$ can be expressed as a Laurent series:  
\begin{equation}
	\A^{-1}(z) = \frac{1}{(z-z_0)^s} \left[ \X_0 + (z-z_0) \X_1 + (z-z_0)^2 \X_2 + \cdots \right],
\end{equation}
where $s$ is a natural number called the order of the pole at $z = z_0$, and $\X_0 \neq 0$.  

To determine the value of $s$ and compute the coefficient matrices $\X_k$, we use the \emph{Sain-Massey}  method~\cite{Sain1969,Howlett1982,Howlett2001}. The method introduces an \emph{augmented matrix}, defined as:
\begin{equation}\label{eq:aug_matrix}
	\bm{\mathcal{A}}^{(t)} := \begin{pmatrix}
		\A_0 & 0      & 0      & \dots & 0\\
		\A_1 & \A_0   & 0      & \dots & 0 \\
		\A_2 & \A_1   & \A_0   & \dots & 0 \\
		\vdots & \vdots & \vdots & \dots & \vdots \\
		\A_t & \A_{t-1} & \A_{t-2} & \dots & \A_0
	\end{pmatrix}.
\end{equation}

The pole order $s$ is identified as the \emph{smallest} value of $t$ for which the rank condition 
\begin{equation}
	\rank \bm{\mathcal{A}}^{(t)} - \rank \bm{\mathcal{A}}^{(t-1)} = n
\end{equation}
is satisfied, where $n$ is the dimension of the matrix $\A(z)$.  

Once $s$ is known, the coefficient matrices $\X_k$ are determined recursively for $k = 1, 2, \dots$ using
\begin{equation}\label{eq:recur}
	\X_k = \sum_{j=0}^{s} \mathcal{G}_{0j}^{(s)} \left( \delta_{j+k,s} \Id - \sum_{i=1}^{k} \A_{i+j} \X_{k-i} \right),
\end{equation}
with the initial value $\X_0$ given by the block $\mathcal{G}_{0s}^{(s)}$, here $\Id$ is the $n$--dimensional  identity matrix.  

Here, $\mathcal{G}_{0s}^{(s)}$ is part of the generalized inverse matrix $\mathcal{G}^{(s)}$, defined as:
\begin{equation}\label{eq:Gmatrix}
	\mathcal{G}^{(s)} := \begin{pmatrix}
		\mathcal{G}_{00}^{(s)} & \dots & \mathcal{G}_{0s}^{(s)} \\
		\vdots & \dots & \vdots \\
		\mathcal{G}_{s0}^{(s)} & \dots & \mathcal{G}_{ss}^{(s)}
	\end{pmatrix},
\end{equation}
which represents the Moore–Penrose pseudo-inverse of the augmented matrix at $t = s$, i.e., $\mathcal{G}^{(s)} = (\bm{\mathcal{A}}^{(s)})^+$ \cite{avrachenkov2013analytic}. Here, the superscript ($+$) denotes the pseudoinverse of a
matrix, which reduces to the standard inverse for a square,
non-singular matrix. This approach systematically determines both the pole order and the expansion coefficients.


	\section{Sain-Massey expansion with the singular dynamical generator given in \eqnref{eq:Hsingular}}\label{eqApp:SMexp}
	
	We now proceed to analyze the Sain–Massey (SM) expansion of the response function at resonance. For clarity, we restate the expansion from \eqnref{eq:SMexpansion} below:
	
	\begin{equation}
		\label{eqAppn:SMexp}
		\smat{G}_\theta[\omega = \omega_0] = \sform (\theta \smat{n} - \smat{H}_\mrm{S})^{-1} = \theta^{-s} \left[\X_0 + \theta \X_1 + \theta^2 \X_2 + \cdots \right],
	\end{equation}
	where the goal is to determine the order $s$ of the pole and the coefficient matrices $\X_\ell$ for a specific class of linear perturbation to mode-frequency in the presence of a singular dynamical generator $\mat{H}_\mrm{S}$ in \eqnref{eq:Hsingular}. Assuming $g = \gamma$, the phase space representation of $\mat{H}_\mrm{S}$, as defined in \eqnref{eq:phase_space_rep}, is given in by the following $8\times8$ matrix:
	\begin{equation}
		\label{eq:Hsing}
		\smat{H}_\mrm{S}|_{g = \gamma} = \begin{pmatrix}
			0 & \gamma & 0 & 0 & \gamma & 0 & 0 & 0 \\
			\gamma & 0 & 0 & 0 & 0 & -\gamma & 0 & 0 \\
			0 & 0 & 0 & \gamma & 0 & 0 & \gamma & 0 \\
			0 & 0 & \gamma & 0 & 0 & 0 & 0 & -\gamma \\
			-\gamma & 0 & 0 & 0 & 0 & \gamma & 0 & 0 \\
			0 & \gamma & 0 & 0 & \gamma & 0 & 0 & 0 \\
			0 & 0 & -\gamma & 0 & 0 & 0 & 0 & \gamma \\
			0 & 0 & 0 & \gamma & 0 & 0 & \gamma & 0
		\end{pmatrix}.
	\end{equation}
	
	We focus on a symmetric four-mode perturbation characterized by the identity matrix
	\begin{equation}
		\label{appn:pert}
		\smat{n} = \Id_{8 \times 8},
	\end{equation}
	which corresponds to uniform modulation of all cavity frequencies~\cite{Zhang2019,Naikoo2023}.
	
	To determine the pole order $s$ and the coefficient matrices $\X_\ell$, we compute the augmented matrices $\bm{\mathcal{A}}^{(j)}$ defined as
	\begin{align}
		\bm{\mathcal{A}}^{(0)} &= \begin{pmatrix} -\smat{H}_\mrm{S}|_{g = \gamma} \end{pmatrix}, \\
		\bm{\mathcal{A}}^{(1)} &= \begin{pmatrix}
			-\smat{H}_\mrm{S}|_{g = \gamma} & 0 \\
			\Id & -\smat{H}_\mrm{S}|_{g = \gamma}
		\end{pmatrix}, \\
		\bm{\mathcal{A}}^{(2)} &= \begin{pmatrix}
			-\smat{H}_\mrm{S}|_{g = \gamma} & 0 & 0 \\
			\Id & -\smat{H}_\mrm{S}|_{g = \gamma} & 0 \\
			0 & \Id & -\smat{H}_\mrm{S}|_{g = \gamma}
		\end{pmatrix}.
		\label{eq:augA2}
	\end{align}
	
	Evaluating the ranks of these matrices, we find
	\begin{equation}
		\rank[\bm{\mathcal{A}}^{(2)}] - \rank[\bm{\mathcal{A}}^{(1)}] = 8,
	\end{equation}
	which equals the dimension of the response matrix $\smat{G}_\theta$. This confirms that the SM expansion has a pole of order $s = 2$.
	
	The next step involves computing the matrix $\mathcal{G}^{(2)} = [\bm{\mathcal{A}}^{(2)}]^+$, the Moore-Penrose pseudoinverse of $\bm{\mathcal{A}}^{(2)}$. We organize $\mathcal{G}^{(2)}$ in block form:
\begin{equation}
	\mathcal{G}^{(2)} = [\bm{\mathcal{A}}^{(2)}]^+ =  
	\left(
	\begin{array}{c|c|c}
		\mathcal{G}_{00}^{(2)}  & 	\mathcal{G}_{01}^{(2)}   &	\mathcal{G}_{02}^{(2)} \\
		\hline
		\cdots&\cdots & \cdots
	\end{array}
	\right)_{24 \times 24},
	\label{eq:MPinv4modes}
\end{equation}
	where the top-right block yields the leading coefficient in the SM expansion:
	\begin{equation}
		\X_0 = \mathcal{G}_{02}^{(2)}.
		\label{eq:X0_4mode}
	\end{equation}
	
	The explicit forms of the relevant blocks are as follows:

		\begin{align}
			\mathcal{G}_{00}^{(2)} &= 
		\gamma\tau\,
		\begin{pmatrix}
			0 & -1 & 0 & 0 &  1 & 0 & 0 & 0\\
			-1 & 0 & 0 & 0 & 0 & -1 & 0 & 0\\
			0 & 0 & 0 & -1 & 0 & 0 &  1 & 0\\
			0 & 0 & -1 & 0 & 0 & 0 & 0 & -1\\
			1 & 0 & 0 & 0 & 0 & -1 & 0 & 0\\
			0 & 1 & 0 & 0 & -1 & 0 & 0 & 0\\
			0 & 0 & 1 & 0 & 0 & 0 & 0 & -1\\
			0 & 0 & 0 & 1 & 0 & 0 & -1 & 0
		\end{pmatrix}, \\
		\mathcal{G}_{01}^{(2)} &=
		\begin{pmatrix}
			\beta & 0      & 0      & 0      & 0      & \alpha & 0      & 0      \\
			0      & \beta & 0      & 0      & -\alpha & 0      & 0      & 0      \\
			0      & 0      & \beta & 0      & 0      & 0      & 0      & \alpha \\
			0      & 0      & 0      & \beta & 0      & 0      & -\alpha & 0      \\
			0      & -\alpha & 0      & 0      & \beta & 0      & 0      & 0      \\
			\alpha & 0      & 0      & 0      & 0      & \beta & 0      & 0      \\
			0      & 0      & 0      & -\alpha & 0      & 0      & \beta & 0      \\
			0      & 0      & \alpha & 0      & 0      & 0      & 0      & \beta
		\end{pmatrix},\\
		\mathcal{G}_{02}^{(2)} &=
		\gamma \,
		\begin{pmatrix}
			0 &  1 & 0 & 0 &  1 & 0 & 0 & 0 \\
			1 &  0 & 0 & 0 &  0 & -1& 0 & 0 \\
			0 &  0 & 0 & 1 &  0 & 0 & 1 & 0 \\
			0 &  0 & 1 & 0 &  0 & 0 & 0 & -1\\
			-1 &  0 & 0 & 0 &  0 & 1 & 0 & 0 \\
			0 &  1 & 0 & 0 &  1 & 0 & 0 & 0 \\
			0 &  0 & -1& 0 &  0 & 0 & 0 & 1 \\
			0 &  0 & 0 & 1 &  0 & 0 & 1 & 0
		\end{pmatrix}
		= \smat{H}_\mathrm{S}\big|_{g = \gamma},
	\end{align}
	where $	\tau \;=\;\frac{1}{4\gamma^2+1}$,  $\alpha \;=\;2\,\gamma^2\,\tau$ and $\beta \;=\;(2\gamma^2+1)\,\tau \;=\;\tau+\alpha$. 
	With these blocks in hand, the remaining coefficients $\X_\ell$ can be recursively determined using the relations:
	\begin{align}
		\X_1 &= -\mathcal{G}_{00}^{(2)} \A_1 \X_0 + \mathcal{G}_{01}^{(2)}, \\
		\X_2 &= \mathcal{G}_{00}^{(2)} (\Id - \A_1 \X_1), \\
		\X_\ell &= -\mathcal{G}_{00}^{(2)} \A_1 \X_{\ell-1} \quad \text{for } \ell \geq 3.
	\end{align}
	Substituting $\A_1 = \Id$ and the expressions above yields $\X_1 = \Id$ ($8 \times 8$ identity matrix) and $\X_\ell = 0$ for $\ell \ge 2$. Hence, the only non-zero coefficients in \eqnref{eqAppn:SMexp} are $\X_1 = \Id$ and $\X_0 = \smat{H}_\mrm{S}|_{g = \gamma}$.   Thus, the complete SM expansion under the symmetric perturbation $\smat{n} = \Id_{8\times8}$ finally reads:
	\begin{equation}
		\smat{G}_{\theta} = \sform \theta^{-2} (\smat{H}|_{g = \gamma} + \theta \Id).
		\label{eq:SME_2mode}
	\end{equation}
	
	The same method can be systematically applied to other perturbation matrices $\smat{n}$.
	
%
\end{document}